\documentclass[preprint,aps,prd,showpacs,nofootinbib]{revtex4}
\usepackage{mathrsfs}
\usepackage{amsmath}
\usepackage{graphicx}
\usepackage{subfigure}

\newcommand{\bea}{\begin{eqnarray}}
\newcommand{\eea}{\end{eqnarray}}
\newcommand{\beq}{\begin{equation}}
\newcommand{\eeq}{\end{equation}}

\def\/{\over}

\begin{document}
\title{\bf Open quantum system approach to Gibbons-Hawking effect of de Sitter space-time}
\author{  Hongwei Yu $^{1,2,}$ }
\affiliation{$^1$ Center for Nonlinear Science and Department of
Physics, Ningbo University, Ningbo, Zhejiang 315211, China\\
$^2$ Institute of Physics and Key Laboratory of Low Dimensional
Quantum Structures and Quantum Control of Ministry of Education,
Hunan Normal University, Changsha, Hunan 410081, China}


\begin{abstract}

We analyze, in the paradigm  of open quantum systems, the reduced
dynamics of a freely falling two-level detector in de Sitter
space-time in weak interaction with a reservoir of fluctuating
quantized  conformal scalar fields in the de Sitter-invariant
vacuum. We find that the detector  is asymptotically driven to a
thermal state at the Gibbons-Hawking temperature, regardless of its
initial state. Our discussion therefore shows that the
Gibbons-Hawking effect of de Sitter space-time can be understood as
a manifestation of thermalization phenomena that involves
decoherence and dissipation in open quantum systems.

\end{abstract}
\maketitle

\baselineskip=16pt

The theory  of quantum open systems has, in recent years, greatly
advanced our understanding of fundamental issues at the foundation
of quantum mechanics and nonequilibrium statistical physics, and it
has been fruitfully applied to the nascent yet fast-growing fields
of quantum information science, modern quantum optics, atomic and
many-body systems, soft condensed matter physics, and biophysics. In
this Letter, we will apply the open quantum system approach to the
investigation of the Gibbons-Hawking effect~\cite{Gibbons} in the
hope of gaining insights into de Sitter space-time from the
prospective of quantum open systems. Our interest in this issue also
lies in the fact that the AdS/CFT correspondence between quantum
gravity on an AdS space-time and a conformal quantum field theory
without gravity on a boundary has provided new and rewarding lines
of research into many branches of physics, such as condensed matter
physics (see Ref.~\cite{Hartnoll09} for a recent review) and quantum
chromodynamics (see Ref.~\cite{Brodsky07} for a recent review), and
there may also exist a holographic duality between quantum gravity
on de Sitter space-time and a conformal field theory living on the
boundary identified with the timelike infinity of de Sitter
space-time~\cite{DSCFT}.

 Our plan is to examine the time evolution of a freely falling two-level detector
 in interaction with fluctuating vacuum conformal
scalar fields in de Sitter space-time. The detector is treated as an
open quantum system and the vacuum with fluctuations of the quantum
fields as the environment. The evolution of the detector is subject
to the effects of decoherence and dissipation due to its interaction
with the environment, and as for any open system, the full dynamics
of the detector can be obtained from the complete time evolution
describing the total system (detector plus the environment) by
integrating over the field degrees of freedom, which are in fact not
observed. We will show that the detector in  de Sitter space-time,
regardless of its initial state, is asymptotically driven to a
thermal state at the Gibbons-Hawking temperature. The open quantum
system approach in our Letter therefore demonstrates that the
Gibbons-Hawking effect can be understood as a manifestation of
thermalization phenomena in the framework of open quantum systems.
It is worth noting here that an examination of similar issues, i.e.,
the Unruh effect and the Hawking radiation, in the paradigm of open
quantum systems has been  carried out, respectively, in
Ref.~\cite{Benatti1} and Ref.~\cite{ZY}

When vacuum fluctuations are concerned in a curved space-time, one
first has to specify the vacuum states.  The vacuum states in de
Sitter space-time can be classified into two categories: one is the
de Sitter-invariant states, the others are those which break de
Sitter invariance~\cite{Allen}. Generally, the de Sitter-invariant
vacuum, whose status in de Sitter space-time is just like Minkowski
vacuum in the flat space-time, is deemed to be a natural vacuum.  So
we will consider the evolution in the proper time  of a freely
falling detector in interaction with a quantized conformally coupled
massless scalar field in the de Sitter-invariant vacuum.

We assume the combined system (detector + external fluctuating
vacuum fields) to be initially prepared in a factorized state, with
the detector having two internal energy levels and the fields in the
de Sitter vacuum. Thus the detector can be fully described in terms
of a two-dimensional Hilbert space, so that its states can be
represented by a $2\times2$ density matrix $\rho$, which is
Hermitian $\rho^+=\rho$, and normalized ${\rm {Tr}}(\rho)=1$ with
${\rm {det}}(\rho)\geq0\;.$
 Without loss of generality, we take the total
Hamiltonian for the complete system  to have the form
 \beq
\label{H}
 H=H_s+H_\phi+ \lambda\;H'\;.
 \eeq
  Here $H_s$ is the Hamiltonian of
the detector, which is taken, for simplicity, to be
 \beq
H_s={\omega_0\over 2}\sigma_3\;,
 \eeq
 where $\sigma_3\;$ is the Pauli
matrix and $\omega_0$ the energy level spacing. $H_\phi$ is the
standard Hamiltonian of conformal scalar fields in de Sitter
space-time, details of which need not be specified here and $H'$ is
the interaction Hamiltonian of the detector with the external scalar
fields and is assumed to be given by
 \beq
  H'= \sigma_{3}\Phi(x)\;.
  \eeq
Let us note here that we can also write  $H_s$ and $H'$ in more
general forms~\cite{Benatti1}, but that does not change the result
of this Letter. In order to achieve a rigorous, mathematically sound
derivation of the reduced dynamics of the detector, we will assume
that the interaction between the detector and the scalar fields is
weak so that the finite-time evolution describing the dynamics of
the detector takes the form of a one-parameter semigroup of
completely positive maps~\cite{Davies,BP}. It should be pointed out
that the coupling constant $\lambda$ in (\ref{H}) should be small,
and this is required by our assumption that the interaction of the
atom with the scalar fields is weak.

 Initially,  the complete system is described by the total density
 matrix
 $\rho_{tot}=\rho(0)\otimes|0\rangle\langle0|\;,$
where $\rho(0)$ is the initial reduced density matrix of the
detector and $|0\rangle$ is the de Sitter-invariant vacuum  state
for field $\Phi(x)$.
  In the frame of the detector,  the evolution in the proper time $\tau $ of the total density
  matrix
$\rho_{tot}$ of the complete system satisfies
\begin{equation}
\frac{\partial\rho_{tot}(\tau)}{\partial\tau}=-iL_H[\rho_{tot}(\tau)]\;,
\label{evolution}
\end{equation}
where the symbol $L_H$ represents the Liouville operator
associated with $H$
\begin{equation}
L_H[S]\equiv[H,S]\;.
\end{equation}
The dynamics of the detector can be obtained by tracing over the
field degrees of freedom, i.e., by applying the trace projection to
the total density matrix $\rho(\tau)={\rm
{Tr}}_{\Phi}[\rho_{tot}(\tau)]\;$.

 In the limit of weak coupling which we assume in this Letter, the reduced density  is found to obey
an equation in the Kossakowski-Lindblad
form~\cite{Lindblad,Benatti0}
\begin{equation}
\frac{\partial\rho({\tau})}{ \partial {{\tau}}}= -i \big[H_{\rm
eff},\, \rho({\tau})\big]
 + {\cal L}[\rho({\tau})]\ ,
\label{master}
\end{equation}
where
\begin{equation}\label{lindblad}
{\cal L}[\rho]=\frac{1}{2} \sum_{i,j=1}^{3} a_{ij}\big[2\,
\sigma_j\rho\,\sigma_i -\sigma_i\sigma_j\, \rho
-\rho\,\sigma_i\sigma_j\big]\;.
\end{equation}
The  matrix $a_{ij}$ and the effective Hamiltonian $H_{\rm eff}$ are
determined by the Fourier transform ${\cal G}(\lambda)$ and Hilbert
transform ${\cal K}(\lambda)$ of the field vacuum correlation
functions (Wightman functions)
\begin{equation}
G^+(x-y)=\langle0|\Phi(x)\Phi(y)|0 \rangle\label{twogreen}\;,
\end{equation}
which are defined as
\begin{equation}
{\cal G}(\lambda)=\int_{-\infty}^{\infty} d\tau \,
e^{i{\lambda}\tau}\, G^{+}\big(x(\tau)\big)\; , \label{fourierG}
\end{equation}
\begin{equation}
{\cal K}(\lambda)= \frac{P}{\pi i}\int_{-\infty}^{\infty} d\omega\
\frac{ {\cal G}(\omega) }{\omega-\lambda} \;. \label{kij}
\end{equation}
Then  the coefficients of the Kossakowski matrix $a_{ij}$ can be
written explicitly as
\begin{equation}
a_{ij}=A\delta_{ij}-iB
\epsilon_{ijk}\delta_{k3}+C\delta_{i3}\delta_{j3}
\end{equation}
with
\begin{equation} \label{ABC}
A=\frac{1}{2}[{\cal {G}}(\omega_0)+{\cal{G}}(-\omega_0)]\;,\;~~~
B=\frac{1}{2}[{\cal {G}}(\omega_0)-{\cal{G}}(-\omega_0)]\;,~~~~
C={\cal{G}}(0)-A\;.
\end{equation}
The effective Hamiltonian $H_{\rm eff}$ contains a correction term,
the so-called Lamb shift, and one can show that it can be obtained
by replacing $\omega_0$ in $H_s$ with a renormalized energy level
spacing $\Omega$ as follows
\begin{equation}
H_{\rm eff}=\frac{\Omega}{2}\sigma_3=\{\omega_0+i[{\cal
K}(-\omega_0)-{\cal K}(\omega_0)]\}\sigma_3\;,
\end{equation}
where a suitable subtraction is assumed in the definition of ${\cal
K}(-\omega_0)-{\cal K}(\omega_0)$ to remove the logarithmic
divergence which would otherwise be present.

In order to find out how the reduced density evolves with proper
time from Eq.~(\ref{master}), we need the Wightman function for the
conformally coupled scalar fields in the de Sitter-invariant vacuum.
Let us note that there are several different coordinate systems that
can be chosen to parametrize  de Sitter space-time~\cite{QFT}. Here
we choose to work with the global coordinate system
$(t,\chi,\theta,\phi)$ in which the freely falling detector is
comoving with the expansion. The line element is
 \beq
ds^2=dt^2-\alpha^2\cosh^2(t/\alpha)[d\chi^2+\sin\chi^2(d\theta^2+\sin^2\theta
     d\varphi^2)]\;
     \label{ds1}
 \eeq
with $\alpha=3^{1/2}\Lambda^{-1/2}$, where $\Lambda$ is the
cosmological constant. The parameter $t$ is often called the world
or cosmic time. The canonical quantization of a massive scalar field
with this metric has been done in Ref.~\cite{QFT,Bunch and
Davies,Tagirov,Ford,
Schomblond,Mottola,Allen,Allen87,Polarski,Polarski prd}. In
coordinates (\ref{ds1}), the wave equation for a massive scalar
field becomes
\begin{eqnarray}
\biggl[{1\/\cosh^3t/\alpha}{\partial\/\partial
t}\biggl({\cosh^3{t\/\alpha}}~{\partial\/\partial
t}\biggr)-{\Delta\/\alpha^2\cosh^2t/\alpha}+m^2+\xi
R\biggr]\phi=0\;,\label{w eq1}
\end{eqnarray}
where the Laplacian
\begin{eqnarray}
\Delta={1\/\sin^2\chi}\biggl[{\partial\/\partial\chi}\biggl(\sin^2\chi{\partial\/\partial\chi}\biggr)
+{1\/\sin\theta}{\partial\/\partial\theta}\biggl(\sin\theta{\partial\/\partial\theta}\biggr)
+{1\/\sin^2\theta}{\partial^2\/\partial\varphi^2}\biggr]\;.
\end{eqnarray}
From (\ref{w eq1}) one can get the eigenmodes, and define a de
Sitter-invariant vacuum. Then the Wightman function can be written
as \cite{Allen87}
\begin{eqnarray}
G^+(x(\tau),x(\tau')))=-{1\/16\pi\alpha^2}~{{1\/4}-\nu^2\/\cos
\pi\nu}~F\biggl({3\/2}+\nu,{3\/2}-\nu;2;{1-Z(x,x')\/2}\biggr)\;,
\end{eqnarray}
where $F$ is a hypergeometric function and
\begin{eqnarray}
&&Z(x,x')=\sinh{t\/\alpha}\sinh{t'\/\alpha}-\cosh{t\/\alpha}\cosh{t'\/\alpha}\cos\Omega\;\nonumber\\&&
\cos\Omega=\cos\chi\cos\chi'+\sin\chi\sin\chi'[\cos\theta\cos\theta'+\sin\theta\sin\theta'\cos(\varphi-\varphi')]'\;,\nonumber\\&&
\nu=\biggl[{9\/4}-{12\/R}(m^2+\xi R)\biggr]^{1/2}\;.
\end{eqnarray}
In the massless, conformal coupling limit, for a freely falling
detector,
the Wightman function becomes
\begin{eqnarray}
G^+(x(\tau),x(\tau')))=-{1\/16\pi^2\alpha^2\sinh^2({\tau-\tau'\/2\alpha}-i\varepsilon)}\;.
\end{eqnarray}
 Its Fourier transform can be found to be
\begin{eqnarray}\label{e4}
{\cal G}_{ds}(\lambda)={\lambda\over 2\pi}\;{
e^{2\pi\alpha\lambda}\over {e^{2\pi\alpha\lambda-1 }}}\;.
\end{eqnarray}
This leads to
 \beq
 \label{AB}
A=1/2[{\cal G}_{ds}(\lambda)+{\cal G}_{ds}(-\lambda)]=\frac{\lambda
\coth(\pi\alpha\lambda)}{4\pi}\;,\;\;\; B=1/2[{\cal
G}_{ds}(\lambda)-{\cal G}_{ds}(-\lambda)]=\frac{\lambda}{4\pi}\;.
 \eeq

In order to solve Eq.(\ref{master}) to find out how the reduced
density evolves with proper time, let us express it in terms of the
Pauli matrices,
\begin{equation}\label{rho-b}
\rho({\tau})=\frac{1}{2}\bigg(1+\sum_{i=1}^{3}\rho_i({\tau})\sigma_i\bigg)\;.
\end{equation}
Substituting Eq.~(\ref{rho-b}) into Eq.~(\ref{master}), one can show
that the Bloch vector $|\rho({\tau})\rangle$ of components
$\{\rho_1({\tau}),\rho_2({\tau}),\rho_3({\tau})\}$ obeys
\begin{equation}\label{sh-eq}
\frac{\partial}{\partial{{\tau}}}|\rho({\tau})\rangle=-2{\cal{H}}|\rho({\tau})\rangle+|\eta\rangle\;,
\end{equation}
where $|\eta\rangle$ denotes a constant  vector $\{0,0,-4B\}$. The
exact form of the matrix ${\cal{H}}$ reads
\begin{equation}
{\cal{H}}=\left(
\begin{array}{ccc}
2A+C& \Omega/2& 0\\ -\Omega/2&2A+C&0 \\ 0&0&2A
\end{array}\right)\;.
\end{equation}
This matrix is nonsingular and its three eigenvalues are
$\lambda_1=2A,\; \lambda_{\pm}=(2A+C)\pm i \Omega /2$. Since the
real parts of these eigenvalues are positive, at later times,
$|\rho({\tau})\rangle$ will reach an equilibrium state
$|\rho({\infty})\rangle$~\cite{Lendi}, which can be found by
formally solving  Eq.~(\ref{sh-eq})
\begin{equation}\label{rhot}
|\rho(\tau)\rangle=e^{-2{\cal{H}}{\tau}}|\rho(0)\rangle+(1-e^{-2{\cal{H}}{\tau}})|\rho_\infty\rangle\;,
\end{equation}
with
 \begin{equation}\label{beta1}
|\rho_\infty\rangle=\frac{1}{2}{\cal{H}}^{-1}|\eta\rangle=-\frac{B}{A}\left(
\begin{array}{c}
 0\\ 0 \\ 1
\end{array}\right)\;.
\end{equation}
If we let $\beta=1/T=2\;{\rm{arctanh}}(B/A)/\omega_0$, we can easily
show that Eq.~(\ref{beta1}) can be rewritten in a purely thermal
form
\begin{equation}\label{beta2}
\rho_\infty=\frac{e^{-\beta {H_s}}}{{\rm{Tr}}[e^{-\beta {H_s}}]}\;.
\end{equation}
Making use of Eq.~(\ref{AB}), we find the temperature of the thermal
state is
 \beq
  T=\frac{\omega}{2{\rm
{arctanh}}(B/A)}={1\over 2\pi\alpha}\;.
 \eeq
 This is exactly the Gibbons-Hawking temperature.
 Thus, regardless of its initial state, a freely falling
two-level detector in de Sitter space-time is asymptotically driven
to a thermal state at the Gibbons-Hawking temperature. Therefore,
there must exist a bath of thermal radiation in de Sitter
space-time. Our open system approach thus reveals that the existence
of the Gibbons-Hawking effect is simply a manifestation of
thermalization phenomena in the framework of open system dynamics.
 At this point, it is worth noting that the Gibbons-Hawking temperature of de
Sitter space-time has also be derived in other different contexts
such as the global embedding approach~\cite{Narnhofer96, Deser} and
the universal Rindler scheme~\cite{Mensky}.

Further aspects of the Gibbons-Hawking effect in terms of the
thermalization phenomena can be studied by examining the  behavior
of the finite-time solution (\ref{rhot}). For this purpose, let us
note that
\begin{equation}\label{expfunction}
e^{-2{\cal{H}}\tau}=\frac{4}{\Omega^2+4C^2}\bigg\{e^{-4A{\tau}}\Lambda_1+2e^{-2(2A+C){\tau}}\bigg[\Lambda_2
\cos(\Omega{{\tau}})+\Lambda_3\frac{\sin(\Omega{{\tau}})}{\Omega}\bigg]\bigg\}\;,
\end{equation}
where
\begin{eqnarray}\label{matrix123}
&&\Lambda_1=\big[(2A+C)^2+\frac{\Omega^2}{4}\big]I-2(2A+C){\cal{H}}+{\cal{H}}^2\;,\nonumber\\
&&\Lambda_2=-2A(A+C)I+(2A+C){\cal{H}}-\frac{1}{2}{\cal{H}}^2\;,\nonumber\\
&&\Lambda_3=2A\big[\frac{\Omega^2}{4}-C(2A+C)\big]I+\big[C(4A+C)-\frac{\Omega^2}{4}\big]{\cal{H}}-C{\cal{H}}^2\;.
\end{eqnarray}
Equation~(\ref{expfunction}) reveals that a  freely falling detector
in de Sitter space-time is subjected to the effects of decoherence
and dissipation characterized by the exponentially decaying factors
involving the real parts of the eigenvalues of ${\cal{H}}$ and
oscillating terms associated with the imaginary part. Therefore, the
Gibbons-Hawking effect as a manifestation of thermalization
phenomena in the framework of open quantum systems actually involves
phenomena of decoherence and dissipation. This suggests that the
vacuum in de Sitter space-time behaves like a fluctuating as well as
a dissipative medium~\cite{Mottola86,Hu95}.  In this regard, our
approach to the derivation of the Gibbons-Hawking effect seems to
shed new light on the issue as compared to other traditional
treatments, and it ties its existence to the effects of decoherence
and dissipation in open quantum systems .

In summary,  we have  analyzed, using the well-known techniques in
the study of open quantum systems, the time evolution of  a freely
falling detector in de Sitter space-time in weak interaction with
fluctuating vacuum  conformal scalar fields in the de
Sitter-invariant vacuum. The detector has been shown to be
asymptotically driven to a thermal state at the Gibbons-Hawking
temperature, regardless of its initial state. Our open system
approach to the issue therefore shows that the Gibbons-Hawking
effect of de Sitter space-time can be understood as a manifestation
of thermalization phenomena in the framework of open quantum
systems, which actually involves the effects of decoherence and
dissipation. It is worthwhile to note that the general techniques
developed in the theory of open quantum systems may also be
applicable to studying other phenomena in curved space-times, such
as particle creation, and may thus provide new insights in the
physical understanding of these phenomena.

\begin{acknowledgments}
We would like to thank the Kavli Institute for Theoretical Physics
China for hospitality where this work was completed during a program
on AdS/CFT. This work was supported in part by the National Natural
Science Foundation of China under Grants No. 11075083 and No.
10935013, the Zhejiang Provincial Natural Science Foundation of
China under Grant No. Z6100077, the K.C. Wong Magna Fund in Ningbo
University, the National Basic Research Program of China under Grant
No. 2010CB832803, and the Program for Changjiang Scholars and
Innovative Research Team in University (PCSIRT,  No. IRT0964).
\end{acknowledgments}

\end{document}